# Observation of Uniform Supercurrent Flow in Polycrystalline K-doped Ba122 by Combined Magneto-optical Imaging and Finite-element Modeling

Shota Ishiwata[1,*], Sunseng Pyon[2], Tsuyoshi Tamegai[2], Mark D. Ainslie[3] and Akiyasu Yamamoto[1]

[1] Department of Biomedical Engineering, Tokyo University of Agriculture and Technology, Koganei, Tokyo 184-8588, Japan
[2] Department of Applied Physics, The University of Tokyo, Bunkyo-ku, Tokyo 113-8656, Japan
[3] Department of Engineering, King's College London, London WC2R 2LS, UK
*Author to whom any correspondence should be addressed

E-mail: s230587z@st.go.tuat.ac.jp

## Abstract

Macroscopic current uniformity in a $(Ba,K)Fe_2As_2$ bulk sample produced by a process that demonstrated high trapped magnetic fields was evaluated through a comparative experimental and modeling approach. The bulk sample, with a well-defined square geometry, exhibited ideal roof-top patterns in magneto-optical (MO) images. Comparison of the magnetic moment, MO images, and finite element modeling results showed good agreement for the critical current density, suggesting that the supercurrent circulates uniformly throughout the sample on the order of MO resolution. These results highlight the importance of enhancing flux pinning strength and microstructural control at the submicron and grain boundary scale in iron-based superconducting polycrystalline materials.

Keywords: Iron-based superconductor, Magneto-optical imaging, FEM modeling, Trapped field magnets, Magnetization behavior

## 1. Introduction

Iron-based superconductors (IBSs) [1,2] exhibit a high critical temperature ($T_c$), second only to cuprates, and have critical fields above 50 T [3,4], combined with low electromagnetic anisotropy [3,5-9]. These properties make them a promising candidate for strong magnet applications. In recent years, various studies on $(Ba,K)Fe_2As_2$ (K-doped Ba122) materials have been reported, and high critical current density ($J_c$) bulks have been successfully produced using high-pressure methods. Bulk materials produced by the spark plasma sintering (SPS) method have been reported exhibiting $J_c$ values of $\sim 1.0 \times 10^5$ A/cm$^2$ (self-field, 5 K) [10], while bulks produced by the hot isostatic pressing (HIP) method have been reported to achieve $\sim 2.3 \times 10^5$ A/cm$^2$ (self-field, 4.2 K) [11]. High $J_c$ values of $\sim 3.8 \times 10^4$ A/cm$^2$ (10 T, 4.2 K) [12] and $\sim 4.9 \times 10^4$ A/cm$^2$ (10 T, 4.2 K) [13] under high magnetic fields have been reported for wires fabricated by the HIP method, and $J_c > 1.0 \times 10^5$ A/cm$^2$ (10 T, 4.2 K) [14,15] for cold- or hot-worked, uniaxially oriented tapes. Demonstration magnets have also been fabricated: bulk magnets by Weiss et al. [16] and Yamamoto et al. [17], and coil magnets by Pyon et al. [13,18] and Ding et al. [19,20], which successfully generated strong magnetic fields of ~1 T.

Foreseeing applications in NMR, accelerators, and other measurement instruments, a key challenge is achieving spatial magnetic field uniformity. To address this, it is essential to enhance both the uniformity and density of the supercurrent. Early studies on the supercurrent uniformity in iron-based polycrystalline materials revealed the following: although IBSs exhibited signs of weak-links, even randomly oriented polycrystalline materials demonstrated macroscopic critical currents that were an order of magnitude higher than those of polycrystalline cuprates [21,22]. However, electromagnetic granularity was also observed in the REFeAs(O,F) (1111 phase) samples [23], indicating supercurrent loops with different length scales and densities. Experiments with bicrystalline thin films showed that the grain boundary transport critical current is not as severely suppressed in IBSs as in $YBa_2Cu_3O_{7-\delta}$, even at high misorientation angles, with a critical grain boundary angle of 5-9° [24,25]. Hysteresis effects have been reported in wires and bulks with large grain sizes, where the magnetization values differ significantly during the application of increasing and decreasing magnetic fields [26,27]. In addition to evaluating these electromagnetic properties, recent studies have reported the characterization of microstructures using electron microscopy on high purity samples [28-33]. Efforts have also been made to analyze microstructures using machine learning and to develop high-throughput processes capable of producing highly characterized samples [17,34-37].

The purpose of this study is to evaluate the microscopic uniformity of the supercurrent at the scale of a few micrometers in a bulk K-doped Ba122 sample that exhibits high $J_c$ properties, fabricated using the high-energy milling method [30,38,39] combined with the SPS method [10,17]. The supercurrent density was evaluated using three different methods: magnetization measurements on a small sample; direct observation of the magnetic flux density distribution by magneto-optical (MO) imaging; and reproduction of the magnetic flux density distribution numerically using the finite element method (FEM). These results showed good agreement, suggesting that the supercurrent circulates uniformly throughout the sample. The limiting factor for the supercurrent is suggested to lie below the scale of approximately a few micrometers, which corresponds to the resolution of MO imaging.

## 2. Experiment

The K-doped Ba122 polycrystalline bulk sample was prepared by SPS using the mechanochemically synthesized precursor powder produced via a high-energy milling method.

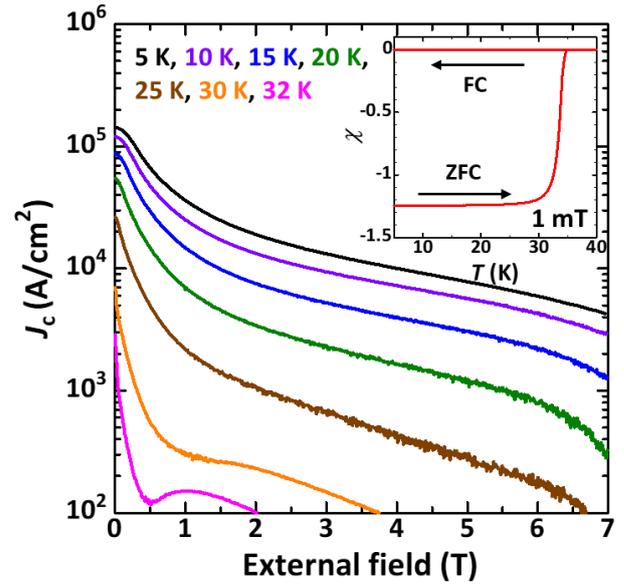

**Figure 1.** External field dependence of critical current density from 5 K to 32 K. The inset shows the temperature dependence of the magnetic susceptibility.

The elemental metals Ba (chunk, 99.9%), K (chunk, 98%), Fe (100 mesh, 99.9+%), and As (granule, 99.9999%) were weighed to achieve a ratio of Ba:K:Fe:As = 0.6:0.4:2:2 in a glove box under an Ar atmosphere. These metals were mixed at high energy [30,38,39] using a planetary ball mill operating at an energy level of 200 MJ/kg to prepare K-doped Ba122 precursor powder. The powder was then filled into a graphite SPS mold with an inner diameter of 10 mm and sintered using an SPS apparatus under uniaxial pressure of 50 MPa, with a heating rate of 50°C/min up to 700°C, followed by cooling to room temperature. The resulting disk-shaped bulk sample had a diameter of 10 mm and a thickness of 1.3 mm. The relative density of the K40%-doped Ba122 sample was 97.5%, with a true density of 5.85 g/cm$^3$ [40]. The grain size of the sample prepared by the same process ranged from tens to hundreds of nanometers [10]. Magnetization measurements were performed on a small specimen (0.52 × 1.48 × 2.41 mm$^3$) cut from the bulk. The temperature dependence of magnetization and the external field dependence of the magnetization at 5 K were measured using a SQUID VSM (MPMS3, Quantum Design). The external field dependence of the magnetization from 10 K to 32 K was measured using the VSM option of the PPMS for a specimen (0.46 × 1.63 × 2.34 mm$^3$) cut from the same batch. $T_c$ was determined as the temperature corresponding to 10% of the magnetic susceptibility at 5 K. The superconducting transition width $\Delta T_c$ was determined by the temperature difference between the points corresponding to 10% and 90% of the magnetic susceptibility at 5 K. $J_c$ was calculated from the magnetization hysteresis loop using the extended Bean model. Fig. 1 shows the external field dependence of $J_c$ at each temperature determined from magnetization measurements, and the inset of Fig. 1 shows the temperature dependence of the magnetic susceptibility. The $J_c$ value at 5 K under self-field was $1.4 \times 10^5$ A/cm$^2$. This value is as high as that of bulk samples [28,41] and wire samples [42] produced by the HIP method. Moreover, at 32 K, a weak supercurrent with a second magnetization peak was observed. The temperature dependence of magnetization (shown in the inset of Fig. 1) reveals $T_c = 34.3$ K with a relatively sharp transition width of $\Delta T_c = 2.7$ K. Since the second magnetization peak appeared near $T_c$, it may have an origin similar to the peak reported in single crystals [6,43].

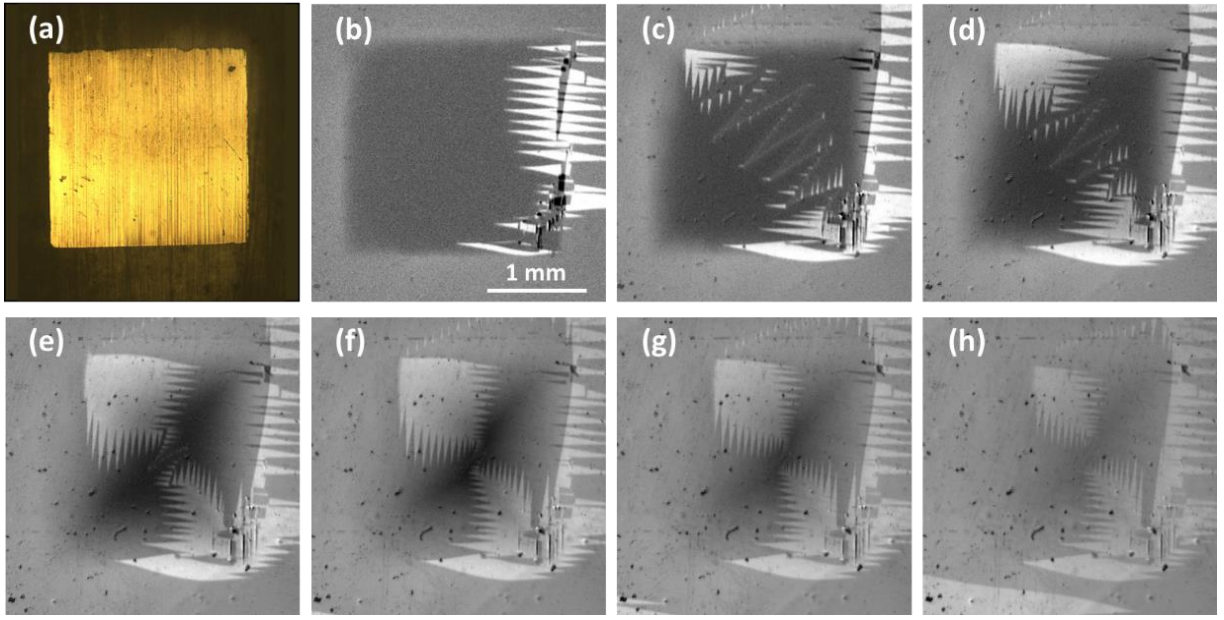

**Figure 2.** (a) Optical microscope image. (b-h) Magneto-optical images of the zero-field cooled bulk sample at 30 K under external fields of (b) 1 mT, (c) 3 mT, (d) 5 mT, (e) 10 mT, (f) 15 mT, (g) 20 mT, and (h) 30 mT.

In the MO imaging, to directly observe the local supercurrent, a regular square prism sample (2.00 × 2.00 × 0.45 mm$^3$) was carefully prepared from the bulk, and the optical contrast of the vertical component of the magnetic flux density ($B_z$) was observed. An iron-garnet indicator film was placed in direct contact with the sample, and the entire assembly was mounted on the cold head of a cryostat (Microstat-HR, Oxford Instruments) and cooled. A cooled CCD camera was used for image acquisition.

The numerical modeling framework is based on the 3D **H**-$\Phi$ formulation [44,45], implemented in the commercial software package COMSOL Multiphysics using the AC/DC module. This mixed formulation significantly reduces the number of degrees of freedom (DOFs), and hence computational time, of the model, whilst maintaining accurate results. To decrease the DOFs further, we exploit the geometric symmetry of the problem to model 1/8$^{th}$ of the entire sample by employing appropriate boundary conditions [45,46]. Thus, half of each dimension (length, width, thickness) is modeled. The $E$-$J$ power law [47,48] is used to simulate the nonlinear resistivity of the superconductor, where $E$ is proportional to $J^n$, and $n$ is the flux creep exponent. Here we assume $n = 35$, which was found to characterize these samples well in [17]. The external field and temperature dependence of the critical current density, $J_c(B, T)$, was input into the model via direct interpolation [49] of the experimental data presented in Fig. 1. To simulate the zero field cooling magnetization process, a ramped magnetic field is applied via magnetic field boundary conditions such that $H_z(t) = H_{app}(t/t_{ramp})$ for $t \leq t_{ramp}$, where $H_{app} = B_{app}/\mu_0$ and $t_{ramp}$ is the duration of the ramp ($t_{ramp}$ = 10 s here). The magnetic field is then ramped down to 0 T at the same ramp rate. Isothermal conditions are assumed; thus, the superconductor is assumed to remain at the same temperature during the magnetization process.

## 3. Results

Fig. 2(a) shows an optical microscope image of the square-shaped sample prepared for MO imaging. Figs. 2(b-h) present MO images captured under external fields ranging from 1 to 30 mT, following cooling to 30 K in the absence of external field. The dark contrast represents regions with low magnetic flux density, while the bright contrast corresponds to regions with high magnetic flux density. The sawtooth-shaped bright contrasts observed in Figs. 2(b-h) result from the in-plane domains of the iron-garnet indicator film. In the image at 1 mT shown in Fig. 2(b), magnetic flux penetration from the edge of the sample was not observed, indicating that the entire sample is in a Meissner state. In the image at 3 mT shown in Fig. 2(c), magnetic flux penetration into the sample is observed. As the applied external field increases, the flux penetrated region gradually expands, as shown in Figs. 2(d, e). In the image at 15 mT shown in Fig. 2(f), the magnetic flux reached the center of the sample. As demonstrated in Figs. 2(b-h), the magnetic flux penetrated approximately uniformly from all sides as the external field is increased. These represent the ideal magnetic flux penetration behavior of a type II superconductor, implying that the shielding current flows uniformly throughout the sample and that macroscopic non-superconducting regions such as defects are negligible.

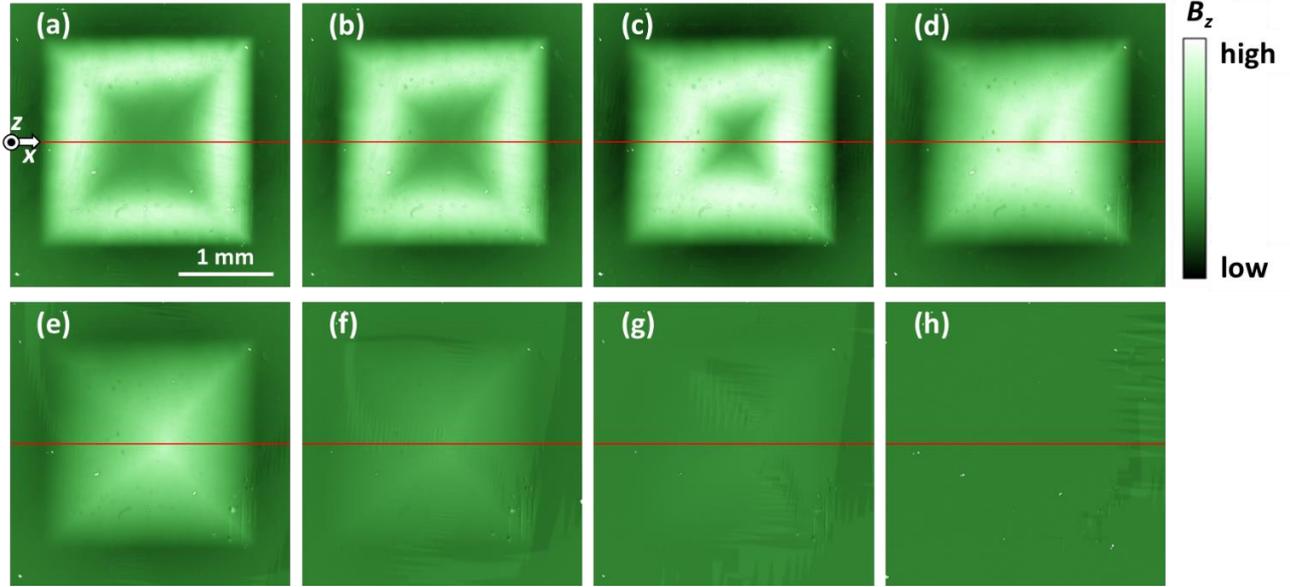

**Figure 3.** (a-h) Magneto-optical images of the remanent magnetic flux density distribution in the sample after zero-field cooling to (a) 5 K, (b) 10 K, (c) 15 K, (d) 20 K, (e) 25 K, (f) 30 K, (g) 32 K, and (h) 40 K, after the application and removal of an external field of 160 mT.

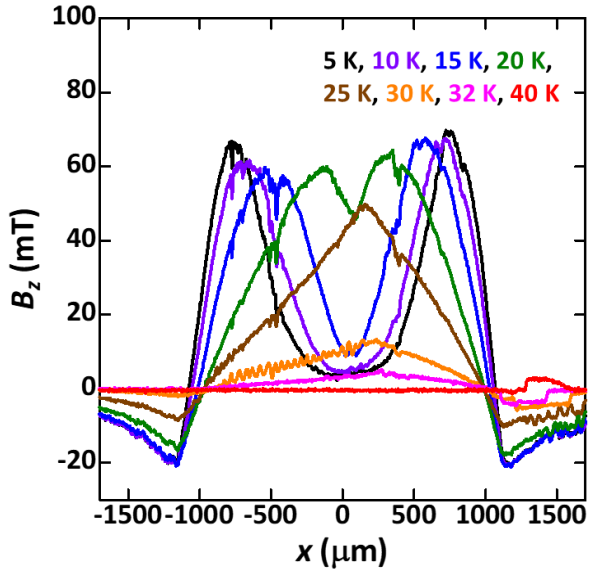

**Figure 4.** Magnetic flux density profiles at each temperature along the red horizontal lines in Fig. 3.

Figs. 3(a-h) show the MO images of the remanent magnetic flux density distribution in the sample after zero-field cooling to predetermined temperatures of 5-40 K, followed by the application and subsequent removal of an external field of 160 mT. Fig. 4 shows the magnetic flux density $B_z(x)$ profile along the $x$-axis, as indicated by the red horizontal lines in Figs. 3(a-h). The magnetic flux density gradient in Fig. 4 is proportional to the local $J_c$ in the sample. At 5 K and 10 K in Figs. 3(a, b), the magnetic flux density remains only at the outer periphery of the sample, while a Meissner (non-penetrated) region exists in the center of the sample. At these temperatures, the bulk interior is shielded from the external field of 160 mT owing to the high $J_c$ of the sample. At 15 K and 20 K in Figs. 3(c, d), the external field reached the center of the sample; however, the sample was not fully magnetized, and roof-top pattern did not appear after the removal of the external field. Roof-top patterns were observed at 25 K and 30 K in Figs. 3(e, f), with the magnetic flux density gradient in Fig. 4 decreasing as $J_c$ decreased with increasing temperature. Surprisingly, even at 32 K, a very weak roof-top pattern was observed in Fig. 3(g). The smooth magnetic flux density gradient shown in Fig. 4 suggests the good homogeneity of $T_c$ within the sample. No

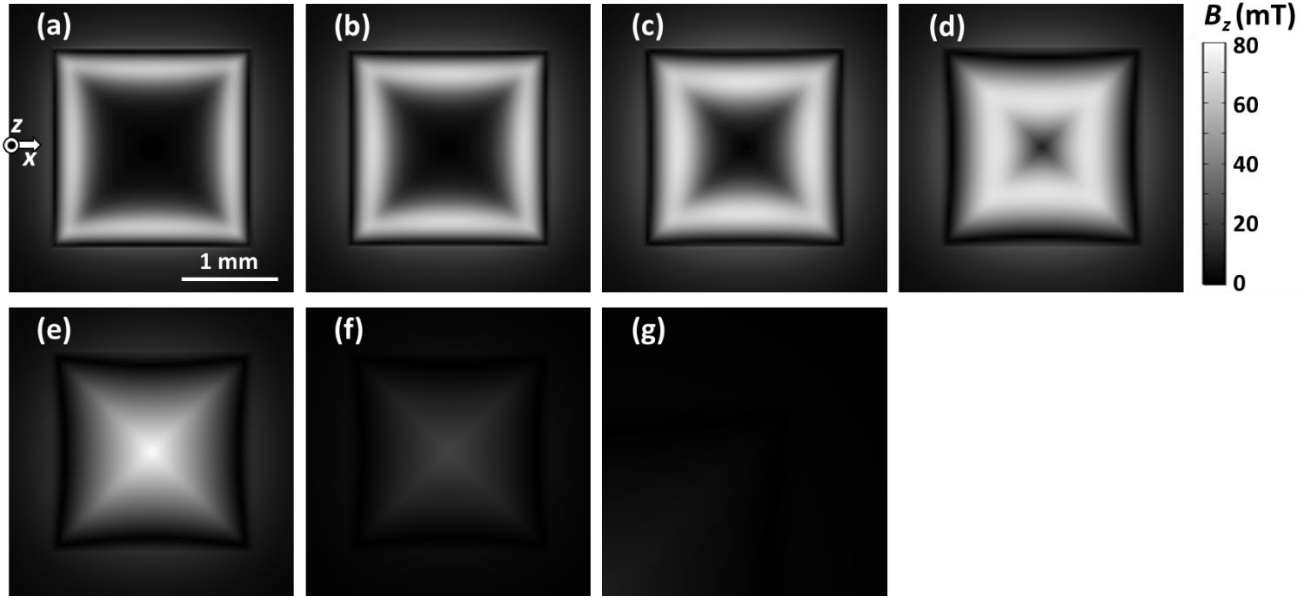

**Figure 5.** (a-g) Simulation results reproducing the MO results shown in Fig. 3 using finite-element modeling. (a) 5 K, (b) 10 K, (c) 15 K, (d) 20 K, (e) 25 K, (f) 30 K, and (g) 32 K.

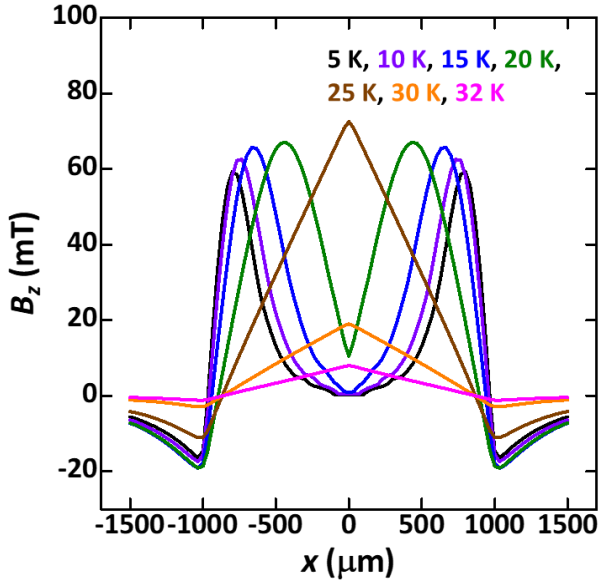

**Figure 6.** Magnetic flux density profiles at each temperature across the center of the surface of sample, as shown in the simulation results of Fig. 5.

signal was observed at 40 K in Fig. 3(h) and Fig. 4, suggesting that there are very few magnetic impurities in the sample at the observation scale. Granularity commonly observed in earlier studies of iron-based polycrystals in the 11 phase [50] and 111 1 phase [21,51-53], was not observed, indicating superior intergranular connectivity in this sample. The magnetic flux density distribution exhibited approximate four-fold symmetry with respect to the sample center at all temperatures, implying uniformly circulating critical current through the entire sample up to near $T_c$ of ~34.3 K.

Two methods were considered to quantitatively evaluate macrocopic $J_c$. In the first method, $J_c$ at 5 K was calculated from the Meissner region width, as shown in Figs. 3(a) and 4 using the following equation [54].

$$J_c = \frac{\pi H_a}{\mathrm{acosh}\left(\frac{w}{b}\right) d} \quad (1)$$

Here, $H_a$ is the external field, $w$ is the width of the sample, $b$ is the width of the Meissner region where the magnetic field does not penetrate, and $d$ is the sample thickness. In Fig. 3(a), the width $b$ of the Meissner region after the application and subsequent removal of 160 mT at 5 K was 1.2 mm, and $J_c = 8.1 \times 10^4$ A/cm$^2$ was estimated. The $J_c$ value under 160 mT at 5 K calculated from magnetic hysteresis loop shown in Fig. 1 was $1.3 \times 10^5$ A/cm$^2$, which is 60% larger than the estimation from equation (1). This discrepancy may be because equation (1) assumes a thin strip as the sample geometry, which differs from the experimental sample geometry.

As the second method, we compared the magnetic flux density distribution experimentally obtained from MO images with that simulated using FEM analysis, which considers the actual sample geometry and external field dependence of $J_c$ (shown in Fig. 1). Fig. 5 shows simulations reproducing the MO images at each temperature shown in Fig. 3. An excellent qualitative agreement between the experimental results and the numerical simulations can be seen, suggesting a uniform supercurrent flow with the assumed $J_c(B)$. Fig. 6 shows magnetic flux density profiles at different temperatures, crossing the center of the sample, as shown in the simulation results of Fig. 5. Experimental and simulated $B_z(x)$ profiles are in close agreement, with some slight discrepancies, suggesting good quantitative agreement of $J_c(B)$ between MO imaging and the simulation assumptions. In particular, the magnetic flux penetrates the superconductor slightly deeper in the experimental case and slightly higher magnetic flux density is calculated (especially for the 25 K case) in the model. This is likely due to the simplification made in the model assuming isothermal conditions; there is likely some small fluctuation in temperature because of the heat generated by magnetic flux movement within the sample during the magnetization process [55]. Such heating would reduce $J_c$ locally, leading to reduced trapped field and hence measured magnetic flux density, as well as increased penetration into the sample.

## 4. Discussion

The MO images of the ZFC samples show clear Meissner regions and roof-top patterns, suggesting that there are no structural defects or weak superconducting regions in the sample that inhibit the current transport and that the supercurrent circulates uniformly. FEM analysis assuming $J_c(B)$, as shown in Fig. 1, shows very good agreement with the experimental MO imaging results. This strongly suggests that the $J_c$ is quantitatively equivalent to the local $J_c$ producing the magnetic flux density gradient observed from the MO images, and that the supercurrent is uniformly circulating throughout the sample. Such uniform supercurrent flow is an advantage of polycrystalline superconducting materials, similar to that reported for $MgB_2$ bulks [56,57].

The reason for the highly uniform superconducting current is due to the fine precursor powder produced by the high-energy milling [30,38,39] and the SPS method [10,17]. The advantage of the short sintering time by the SPS method, in addition to suppressing coarse grain growth, is that it minimizes compositional changes resulting from the evaporation of volatile potassium. Noteworthy is the MO imaging at 30 K, which is close to $T_c$ ~34 K. Fig. 2(b) shows the MO image of the state corresponding to 30 K on the ZFC curve of the inset of Fig. 1. Though 30 K is the temperature just below the superconducting transition, Fig. 2(b) shows the Meissner state of the entire sample. Moreover, in Figs. 2(b-h) and Fig. 3(f), macroscopic circulations of supercurrent are observed, indicating that the sample does not have low $T_c$ regions.

Compared to the $J_c$ value in this study for a K-doped Ba122 polycrystalline bulk, 1-2 orders of magnitude higher $J_c$ values have been reported for single crystals and thin films [6,43,58-61]. This suggests that the $J_c$ is uniformly suppressed throughout the present sample. Such critical current limitation is considered to occur on a scale of less than about a few micrometers, the resolution of MO imaging, and mesoscopic microstructures and grain boundaries can be the specific factors. Recently, Hatano *et al.* evaluated the relationship between the intergranular current transport properties and misorientation angle in K-doped Ba122 bicrystal thin films and reported excellent $J_c$ values exceeding $10^5$ A/cm$^2$ at 28 K for a 24° artificial grain boundary [62]. Untextured polycrystalline samples such as those in this study should contain natural grain boundaries with misorientation angles of more than 24°. Therefore, elucidation of the microscopic grain boundary transport mechanism in K-doped Ba122 through the fabrication and evaluation of artificial single grain boundaries including high misorientation angle is necessary, as well as the understanding of the basic flux pinning mechanism. As for processing, research and development of effective methods for mesoscopic microstructure control and enhancement of flux pinning to improve critical current properties is required. As an illustration of the limitations of this experiment, the specifications of the MO indicator film constrain the magnetic field to a maximum of 80 mT. Consequently, the behavior of the supercurrent at higher magnetic fields, where flux pinning becomes more prominent, remains unexplored in this study. Future research should focus on evaluating the uniformity and dynamics of the supercurrent in the presence of stronger magnetic fields, as well as the impact of enhanced flux pinning, to gain a more comprehensive understanding of the performance of polycrystalline K-doped Ba122.

## 5. Summary

The uniformity of the supercurrent in K-doped Ba122 polycrystalline bulk prepared by high-energy milling and SPS method was evaluated using MO imaging and FEM analysis. The MO imaging of the well-defined, square-shaped sample showed that magnetic flux penetrates uniformly even at 30 K which is close to $T_c$, and the remanent flux density distribution exhibits four-fold symmetry. The remanent flux density distributions observed in MO imaging showed excellent agreement with FEM simulation, suggesting that the supercurrent circulates macroscopically and uniformly with a density approximately equal to the local $J_c$ value. The observed uniform current flow, even in randomly oriented polycrystalline Ba122 with intrinsic weak-links, is promising for the generation of highly uniform magnetic fields with iron-based high temperature superconducting magnets.

## Acknowledgements

This work was supported by JST CREST (JPMJCR18J4), JSPS KAKENHI (JP21H01615), and the Collaborative Research Projects of Laboratory for Materials and Structures, Institute of Innovative Research, Tokyo Institute of Technology.

**References**

[1] Kamihara Y, Watanabe T, Hirano M and Hosono H 2008 *J. Am. Chem. Soc.* **130** 3296
[2] Hosono H, Yamamoto A, Hiramatsu H and Ma Y 2018 *Mater. Today* **21** 278
[3] Tarantini C, Gurevich A, Jaroszynski J, Balakirev F, Bellingeri E, Pallecchi I, Ferdeghini C, Shen B, Wen H H and Larbalestier D C 2011 *Phys. Rev. B* **84** 184522
[4] Altarawneh M M, Collar K, Mielke C H, Ni N, Bud'ko S L, and Canfield P C 2008 *Phys. Rev. B* **78** 220505
[5] Hänisch J, Iida K, Kurth F, Reich E, Tarantini C, Jaroszynski J, Förster T, Fuchs G, Hühne R, Grinenko V, Schultz L and Holzapfel B 2015 *Scientific Reports* **5** 17363
[6] Ishida S, Song D, Ogino H, Iyo A, Eisaki H, Nakajima M, Shimoyama J and Eisterer M 2017 *Phys. Rev. B* **95** 014517
[7] Vinod K, Satya A T, Sharma S, Sundar C S and Bharathi A 2011 *Phys. Rev. B* **84** 012502
[8] Yamamoto A, Jaroszynski J, Tarantini C, Balicas L, Jiang J, Gurevich A, Larbalestier D C, Jin R, Sefat A S, McGuire M A, Sales B C, Christen D K and Mandrus D 2009 *Appl. Phys. Lett.* **94** 062511
[9] Yuan H Q, Singleton J, Balakirev F F, Baily S A, Chen G F, Luo J L and Wang N L 2009 *Nature* **457** 565
[10] Tokuta S, Hasegawa Y, Shimada Y and Yamamoto A 2022 *iScience* **25** 103992
[11] Pak C, Su Y F, Collantes Y, Tarantini C, Hellstrom E E, Larbalestier D C and Kametani F 2020 *Supercond. Sci. Technol.* **33** 084010
[12] Pyon S, Suwa T, Tamegai T, Takano K, Kajitani H, Koizumi N, Awaji S, Zhou N and Shi Z 2018 *Supercond. Sci. Technol.* **31** 055016
[13] Pyon S, Ito T, Tamegai T, Kajitani H, Koizumi N, Awaji S, Kito H, Ishida S and Yoshida Y 2022 *J. Phys.: Conf. Ser.* **2323** 012020
[14] Huang H, Yao C, Dong C, Zhang X, Wang D, Cheng Z, Li J, Awaji S, Wen H H and Ma Y 2018 *Supercond. Sci. Technol.* **31** 015017
[15] Gao Z, Togano K, Zhang Y, Matsumoto A, Kikuchi A and Kumakura H 2017 *Supercond. Sci. Technol.* **30** 095012
[16] Weiss J D, Yamamoto A, Polyanskii A A, Richardson R B, Larbalestier D C and Hellstrom E E 2015 *Supercond. Sci. Technol.* **28** 112001
[17] Yamamoto A, Tokuta S, Ishii A, Yamanaka A, Shimada Y and Ainslie M D 2024 *NPG Asia Mater.* **16** 29
[18] Pyon S, Ito T, Sakagami R, Tamegai T, Awaji S, Kito H, Ishida S, Eisaki H, Yoshida Y and Kajitani H 2023 *Supercond. Sci. Technol.* **36** 015009
[19] Ding H, Zhao H, Huang P, Yu L, Xu J, Fang Z, Chen Z, Wang D, Zhang X and Chen W 2023 *Supercond. Sci. Technol.* **36** 11LT01
[20] Dong C H, Xu Q J, Ma Y W 2024 *Nati. Sci. Rev.* **11** nwae122
[21] Yamamoto A, Polyanskii A A, Jiang J, Kametani F, Tarantini C, Hunte F, Jaroszynski J, Hellstrom E E, Lee P J, Gurevich A, Larbalestier D C, Ren Z A, Yang J, Dong X L, Lu W and Zhao Z X 2008 *Supercond. Sci. Technol.* **21** 095008
[22] Palenzona A, Sala A, Bernini C, Braccini V, Cimberle M R, Ferdeghini C, Lamura G, Martinelli A, Pallecchi I and Romano G 2012 *Supercond. Sci. Technol.* **25** 115018
[23] Yamamoto A, Jiang J, Kametani F, Polyanskii A, Hellstrom E, Larbalestier D, Martinelli A, Palenzona A, Tropeano M and Putti M 2011 *Supercond. Sci. Technol.* **24** 045010
[24] Lee S, Jiang J, Weiss J D, Folkman C M, Bark C W, Tarantini C, Xu A, Abraimov D, Polyanskii A, Nelson C T, Zhang Y, Baek S H, Jang H W, Yamamoto A, Kametani F, Pan X Q, Hellstrom E E, Gurevich A, Eom C B and Larbalestier D C 2009 *Appl. Phys. Lett.* **95** 212505
[25] Katase T, Ishimaru Y, Tsukamoto A, Hiramatsu H, Kamiya T, Tanabe K and Hosono H 2011 *Nat. Commun.* **2** 409
[26] Luo J Y, Okada T, Awaji S, Liu C and Ma Y W 2023 *IEEE Trans. Appl. Supercond.* **33** 8200405
[27] Hecher J, Baumgartner T, Weiss J D, Tarantini C, Yamamoto A, Jiang J, Hellstrom E E, Larbalestier D C and Eisterer M 2016 *Supercond. Sci. Technol.* **29** 025004
[28] Kametani F, Su Y F, Collantes Y, Pak C, Tarantini C, Larbalestier D C and Hellstrom E E 2020 *Appl. Phys. Express* **13** 113002
[29] Shimada Y, Tokuta S, Yamanaka A, Yamamoto A and Konno T J 2022 *J. Alloy. Compd.* **923** 166358
[30] Guo Z M, Muraoka K, Gao H Y, Shimada Y, Harada T, Tokuta S, Hasegawa Y, Yamamoto A and Hata S 2024 *Acta Mater.* **266** 119648
[31] Liu S F, Yao C, Huang H, Dong C H, Guo W W, Cheng Z, Zhu Y C, Awaji S and Ma Y W 2021 *Sci. China Mater.* **64** 2530
[32] Kametani F, Su Y, Tarantini C, Hellstrom E, Matsumoto A, Kumakura H, Togano K, Huang H and Ma Y 2024 *Appl. Phys. Express* **17** 013004
[33] Tarantini C, Pak C, Su Y F, Hellstrom E E, Larbalestier D C and Kametani F 2021 *Scientific Reports* **11** 3143
[34] Hirabayashi Y, Iga H, Ogawa H, Tokuta S, Shimada Y and Yamamoto A 2024 *NPG Comput. Mater.* **10** 46
[35] Ishii A, Kikuchi S, Yamanaka A and Yamamoto A 2023 *J. Alloy. Compd.* **966** 171613
[36] Bagni T, Bovone G, Rack A, Mauro D, Barth C, Matera D, Buta F and Senatore C 2021 *Scientific Reports* **11** 7767
[37] Yamamoto A, Yamanaka A, Iida K, Shimada Y and Hata S 2024 *Sci. Technol. Adv. Mater.* **26** 2436347
[38] Tokuta S and Yamamoto A 2019 *APL Mater.* **7** 111107
[39] Tokuta S, Shimada Y and Yamamoto A 2020 *Supercond. Sci. Technol.* **33** 094010
[40] Rotter M, Pangerl M, Tegel M and Johrendt D 2008 *Angew. Chem. Int. Ed.* **47** 7949
[41] Nikolo M, Weiss J D, Singleton J, Jiang J and Hellstrom E E 2018 *IEEE Trans. Appl. Supercond.* **28** 7300104
[42] Tamegai T, Pyon S, Tsuchiya Y, Inoue H, Koizumi N and Kajitani H 2015 *IEEE Trans. Appl. Supercond.* **25** 7300504
[43] Yang H, Luo H Q, Wang Z S and Wen H H 2008 *Appl. Phys. Lett.* **93**, 142506
[44] Arsenault A, Sirois F and Grilli F 2021 *IEEE Trans. Appl. Supercond.* **31** 680011
[45] Zhang K, Ainslie M, Calvi M, Kinjo R and Schmidt T 2021 *Supercond. Sci. Technol.* **34** 094002
[46] Ainslie M D, Huang K Y, Fujishiro H, Chaddock J, Takahashi K, Namba S, Cardwell D A and Durrell J H 2019 *Supercond. Sci. Technol.* **32** 034002
[47] Rhyner J 1993 *Physica* C **212** 292

[48] Plummer C J G and Evetts J E 1987 *IEEE Trans. Magn.* **23** 1179
[49] Ainslie M D, Fujishiro H, Mochizuki H, Takahashi K, Shi Y-H, Namburi D K, Zou J, Zhou D, Dennis A R and Cardwell D A 2016 *Supercond. Sci. Technol.* **29** 074003
[50] Ding Q P, Mohan S, Tsuchiya Y, Taen T, Nakajima Y and Tamegai T 2011 *Supercond. Sci. Technol.* **24** 075025
[51] Prozorov R, Tillman M E, Mun E D and Canfield P C 2009 *New J. Phys.* **11** 035004
[52] Kametani F, Polyanskii A A, Yamamoto A, Jiang J, Hellstrom E E, Gurevich A, Larbalestier D C, Ren Z A, Yang J, Dong X L, Lu W and Zhao Z X 2009 *Supercond. Sci. Technol.* **22** 015010
[53] Tamegai T, Nakajima Y, Tsuchiya Y, Iyo A, Miyazawa K, Shirage P M, Kito H and Eisaki H 2009 *Physica C* **469** 915
[54] Brandt E H and Indenbom M 1993 *Phys. Rev. B* **48** 12893
[55] Ainslie M and Fujishiro H 2019 *Numerical modelling of bulk superconductor magnetisation* (IOP Publishing)
[56] Larbalestier D C, Cooley L D, Rikel M O, Polyanskii A A, Jiang J, Patnaik S, Cai X Y, Feldmann D M, Gurevich A, Squitieri A A, Naus M T, Eom C B, Hellstrom E E, Cava R J, Regan K A, Rogado N, Hayward M A, He T, Slusky J S, Khalifah P, Inumaru K and Haas M 2001 *Nature* **410** 186
[57] Yamamoto A, Ishihara A, Tomita M and Kishio K 2014 *Appl. Phys. Lett.* **105**, 032601
[58] Iida K, Qin D Y, Tarantini C, Hatano T, Wang C, Guo Z M, Gao H Y, Saito H, Hata S, Naito M and Yamamoto A 2021 *NPG Asia Mater.* **13** 68
[59] Qin D, Iida K, Guo Z, Wang C, Saito H, Hata S, Naito M and Yamamoto A 2022 *Supercond. Sci. Technol.* **35** 09LT01
[60] Qin D Y, Guo Z M, Tarantini C, Hata S, Naito M and Yamamoto A 2024 *Appl. Phys. Lett.* **125** 182601
[61] Taen T, Ohtake F, Pyon S, Tamegai T and Kitamura H 2015 *Supercond. Sci. Technol.* **28** 085003
[62] Hatano T, Qin D Y, Iida K, Gao H Y, Guo Z M, Saito H, Hata S, Shimada Y, Naito M and Yamamoto A 2024 *NPG Asia Mater.* **16** 41